\documentclass[]{pasj01}
\draft

\begin{document} 
\Received{}
\Accepted{}

\title{High-mass star formation in Orion triggered by cloud-cloud collision II; Two merging molecular clouds in NGC2024}

\author{Akio \textsc{Ohama}\altaffilmark{1}, Daichi \textsc{Tsutsumi}\altaffilmark{1}, Hidetoshi \textsc{Sano}\altaffilmark{1},  Kazufumi \textsc{Torii}\altaffilmark{2}, Atsushi \textsc{Nishimura}\altaffilmark{1}, Kazuhiro \textsc{Shima}\altaffilmark{3}, Habe \textsc{Asao}\altaffilmark{3} Hiroaki \textsc{Yamamoto}\altaffilmark{1}, Kengo \textsc{Tachihara}\altaffilmark{1} Yutaka \textsc{Hasagawa}\altaffilmark{4}, Kimihiro \textsc{Kimura}\altaffilmark{4}, Hideo \textsc{Ogawa}\altaffilmark{4} and Yasuo \textsc{Fukui}\altaffilmark{1}}
\altaffiltext{1}{Department of Physics, Nagoya University, Furo-cho Chikusa-ku Nagoya, 464-8602, Japan}
\altaffiltext{2}{Nobeyama Radio Observatory, 462-2 Nobeyama, Minamimaki Minamisaku, Nagano, 384-1305, Japan}
\altaffiltext{3}{Faculty of Science, Department of Physics, Hokkaido University, Kita 10 Nishi 8 Kita-ku,Sapporo 060-0810, Japan}
\altaffiltext{4}{Department of Physical Science, Graduate School of Science, Osaka Prefecture University,}
\email{ohama@a.phys.nagoya-u.ac.jp}



\maketitle

\begin{abstract}
 We analyzed the NANTEN2 $^{13}$CO ($J$=2--1 and 1--0) datasets in NGC 2024. We found that the cloud consists of two velocity components, whereas the cloud shows mostly single-peaked CO profiles. The two components are physically connected to the H{\sc ii}  region as evidenced by their close correlation with the dark lanes and the emission nebulosity. The two components show complementary distribution with a displacement of 0.4 pc. Such complementary distribution is typical to colliding clouds discovered in regions of high-mass star formation. We hypothesize that cloud-cloud collision between the two components triggered the formation of the late O stars and early B stars localized within 0.3 pc of the cloud peak. The collision timescale is estimated to be $\sim$ 10$^5$ yrs from a ratio of the displacement and the relative velocity 3--4 km s$^{-1}$ corrected for probable projection. The high column density of the colliding cloud 10$^{23}$ cm$^{-2}$ is similar to those in the other massive star clusters in RCW 38, Westerlund 2, NGC 3603, and M42, which are likely formed under trigger by cloud-cloud collision. The present results provide an additional piece of evidence favorable to high-mass star formation by a major cloud-cloud collision in Orion.
\end{abstract}

\section{Introduction}
 The Orion region is the most outstanding high-mass star forming region in the solar neighborhood. The H{\sc ii} region NGC 2024 
 located at 410 pc
  is the second active H{\sc ii} region next to M42
  in Orion
  and is associated with 
  a reflection nebula NGC 2023 (e.g., Anthony-Twarog 1982; Menten et al. 2007) 
 . NGC 2024 is therefore one of the most interesting regions in studying high-mass star formation. A review of the Orion B region is given by Meyer et al.(2008; see also references therein)

Observations in radio continuum radiation and recombination lines \citep{kru82, bar89,bik03,ron03} as well as infrared observations indicate that 
NGC 2024
 is ionized by exciting stars of O8V-B2V \citep{lad91,com96,gia00,hai00,bik03,kan07}. In addition to the high-mass stars, low-mass stars are identified
 in the cluster
 by near-infrared and X ray observations \citep{lad91,ski03}. These
 cluster members
 consist of at least 300 stars including late O stars and early B stars. Most of them are suggested to be in mass accretion phase by 10 $\mu$m observations \citep{hai00,hai01}. Large visual extinction of 30 magnitudes toward the region (e.g., Johnstone et al.(2006)) 
 hampers firmly identifying candidates for ionizing stars of NGC 2024. It is probable that the cluster is smaller than the Orion Nebula Cluster which contains $\sim$10 O / early B stars as well as $\sim$2000 member stars \citep{kro01,ban15}. Near infrared observations
 of NGC 2024 was used to construct an HR diagram of these stars and the age is estimated to be 1 Myrs or less \citep{get14}. Far infrared and sub-mm observations revealed dense protostellar condensations with a compact distribution, suggesting some external triggering to form them while details remain elusive due to high extinction.

\citet{fuk17b} made a detailed analysis of the $^{12}$CO ($J$=1--0)  data in the Orion A cloud taken with the 45 m telescope by \citet{shi11}, and presented an analysis that the Orion A cloud comprises two spatially overlapping components of different velocities. These authors suggested that the ionizing stars of M42 and M43 in Orion A were formed by triggering in collision between two clouds of 7 km s$^{-1}$ velocity difference based on that the two clouds show complementary distribution with a systematic spatial displacement, a typical signature of colliding clouds, for M43. If a similar scenario is applicable to the other regions of high-mass star formation, the Orion B cloud is an obvious target to search for a signature of cloud-cloud collision in Orion. In the Orion A cloud we found collision signatures over a 1 degree, which is significantly larger than the typical observed areas of high resolution CO observations in the Orion B cloud; in NGC 2024, the JCMT covers in CO ($J$=3--2) only $10.8 \times 22.5$ arcmin$^2$ \citep{buc10} and a large-scale view of the molecular gas was not revealed.

The Orion B cloud was not a subject of many large-scale molecular observations. \citet{bal91} observed the cloud in the $^{13}$CO $J$=1--0 transition and \citet{aoy01} in the C$^{18}$O $J$=1--0 and HCO$^+$ $J$=1--0 transitions. \citet{rip13}  used the FCRAO 14 m telescope to map the cloud in the $^{12}$CO $J$=1--0 transition at higher resolution. The $^{12}$CO $J$=2--1 transition was observed at 9 arcmin resolution in a large scale by \citet{sak94} and \citet{wil05} and used to derive density and temperature through a comparison with the $^{12}$CO $J$=1--0 transition. Recently, \citet{nis15}  made a higher resolution comparative study of the $^{12}$CO $J$=1--0 and $J$=2--1 transitions by using the data taken with the OPU 1.85 m and NANTEN, and derived density and temperature distributions over the whole Orion region including Orion A and Orion B at 3 arcmin.

In order to reveal detailed gas kinematics in NGC 2024, we carried out new observations toward NGC 2024 cloud in the $^{12}$CO and $^{13}$CO $J$=2--1  transitions at 1.5 arcmin resolution over a large area. Section 2 gives details of the observations, Section 3 describes the observational results, Section 4 presents discussion on cloud-cloud collision, and Section 5 concludes the paper.

\section{Observations}
Observations of the $^{13}$CO ($J$=2--1) transition were made with NANTEN2 over an area of 0.5 degree $\times$ 0.5 degree in $l$ and $b$. The data were taken in a period from December 11, 2016 to December 15, 2016. The transition was observed simultaneously with the $^{12}$CO ($J$=2-1) emission in the on-the-fly mode in 0.8 second integration per a point with a 30 arcsec grid spacing. The system noise temperature in the Single Side Band was 320 - 440 K toward the zenith. The final noise fluctuations were 0.45 - 0.58 K/ch. The backend was a digital spectrometer having a band width and resolution of 1 GHz and 61 kHz, respectively. These correspond to a velocity coverage of 2600 km s$^{-1}$ and a velocity resolution of 0.17 km s$^{-1}$. Pointing accuracy was measured toward IRC 10216 (R.A., Dec.) = (9h47m57s.406, 13$^{\circ}$ 16' 43$\arcsec$.56) everyday and confirmed to be better than 10 arcsec. The absolute intensity scale was established by observing OriKL (R.A., Dec.) = (5h35m13s.5, -5$^{\circ}$ 22' 27$\arcsec$.6) every hour. We use the Galactic coordinate in the present paper.

\section{Results}
\subsection{$^{13}$CO distribution}
Figures 1a and 1b show distributions of the integrated intensity of the $^{13}$CO ($J$=2--1) emission in the NGC 2023 and NGC 2024 region
 at 3 arcmin resolution obtained with the OPU 1.85 m telescope \citep{nis15}. The distributions show the cloud for two velocity ranges, 8.7 - 9.4 km s$^{-1}$ and 11.2 - 11.8 km s$^{-1}$. Figure 1c is a longitude-velocity diagram integrated in latitude. The results indicate that the $^{13}$CO velocity varies significantly in the cloud in a range from 8 to 12 km s$^{-1}$. The $^{13}$CO emission in NGC 2024 shows an extent similar to the optical H{\sc ii} region, suggesting coexistence of molecular and H{\sc ii} gas (Figure 1a). 
NGC 2023 and NGC 2024 are located toward intense CO-emitting regions with broad velocity components.

Figure 2 shows typical $^{12}$CO and $^{13}$CO ($J$=2--1) profiles in two positions P1 and P2 of NGC 2024. The $^{12}$CO emission is heavily saturated with self-absorption in Figure 2a, while $^{13}$CO  having two peaks seems to be non-saturated. We used only $^{13}$CO which traces better gas distribution in the present paper. Positions P1 and P2 are explained further below.

Figure 3 shows the velocity channel distribution of the $^{13}$CO($J$=2--1) emission with NANTEN2 every 0.5 km s$^{-1}$ at a 60 arcsec grid spacing. We found that the primary peak P1 in 10.6 - 11.2 km s$^{-1}$ at ($l$,$b$)=(206.5$^{\circ}$, -16.35$^{\circ}$) and the secondary peak P2 in 8.4 - 9.0 km s$^{-1}$ at ($l$,$b$)=(206.37$^{\circ}$,-16.42$^{\circ}$). The two positions show integrated $^{13}$CO intensity greater than 8 K km s$^{-1}$ in Figure 3. We also found a marked intensity depression D1 in 8 - 10 km s$^{-1}$ peaked at ($l$,$b$)=(206.47$^{\circ}$, -16.4$^{\circ}$).

The DSS2 image shows a major optical dark lane in the H{\sc ii} region. Some of the additional minor dark lanes show a tilt of $\sim$ 70 degrees to the major lane. In order to show detailed correspondence between the dark lanes with molecular gas Figure 4 presents overlays of $^{13}$CO on a DSS2 image. Figures 4a-e present $^{13}$CO features for five velocity ranges superposed on the DSS2 image, and Figure 4f summarize seven dark lanes. Figure 4g illustrates the velocity ranges for the $^{13}$CO features toward each dark lane. The blue-shifted gas shows better correlation with the dark lanes, suggesting that the blue-shifted gas is located on the near side of the H{\sc ii} region. The red-shifted gas lies inside or behind the H{\sc ii} region lanes as shown by its poor correlation with the dark lanes (Figures 4d and 4e).

\subsection{Complementary distribution}
Figure 3 indicates that P1 is pronounced in the cloud and is associated with the infrared cluster including IRS2b (Section 4.3). D1 is an unusual marked depression with sharp intensity gradient from its surroundings. redThere is no known stellar cluster inside D1 (Section 4.3). If cloud-cloud collision is operating, we expect some complementary distribution in Figure 3. Figure 5 shows integrated intensity distributions in two velocity ranges at 8.5 - 10.0 km s$^{-1}$ and 11.2 - 11.7 km s$^{-1}$, where we see two peaks P 1 and P2, and an intensity depression D1. By eye inspection, we recognized complementarity between the intensity peak P1 in Figure 5b and the intensity depression D1 at ($l$,$b$)=(206.46$^{\circ}$, -16.40$^{\circ}$) in Figure 5a, whereas there is a spatial gap of $\sim$0.1 degrees between them. Figure 6 shows that the peak coincides well with the depression if a displacement is applied as indicated by an arrow. 
Such a displacement is an observational signature of colliding clouds 
\citep{fuk17b,fuk18b}.
We used the overlapping integral method \citep{fuk17b} to derive a displacement which fits best the intensity peak P1 with the depression D1 in complementary distribution, where we adopted a new Cartesian coordinate system R and S in order to indicate a relative position from P1 where the overlapping between P1 and D1 becomes maximum (Figure 7). As a result, we estimated an optimum displacement to be R = 0.4 pc at a position angle of 60 deg. (Figure 7). The displacement fits P1 with the intensity depression D1 in the blue-shifted gas, and P2 with the weakly-emitting area toward ($l$, $b$)=(206.45$^{\circ}$, -16.35$^{\circ}$) in the red-shifted gas on the west of P1.
 We also note that the red-shifted component traces the northern edge of P1 in Figure 6b.

\section{Discussion; cloud-cloud collision which triggered the formation of the high-mass stars in NGC 2024}
\subsection{The complementary distribution and the cloud-cloud collision scenario}
We analyzed the $^{13}$CO distribution in NGC 2024. We found a marked intensity peak P1 toward IRS2b and the infrared cluster and a pronounced intensity depression D1. P1 is most likely the star forming dense clump. D1 has no stars inside which can produce the cavity by stellar feedback, and its sharp intensity gradient is unusual as a cloud distribution which originated in a self-gravitational process. We found that the cloud comprises two velocity components with complementary distribution, where the projected velocity separation is $\sim$2 km s$^{-1}$. The peak P1 is in the red-shifted component and the intensity depression D1 in the blue-shifted component. These properties suggest that P1 and D1 are related via cloud-cloud collision process which has been discussed for more than 20 objects in the literature (Section 1). According to the theoretical simulations of \citet{tak14} and the synthetic observations by \citet{fuk17b}, an intensity depression is produced by a cloud-cloud collision where a small cloud creates a cavity in a large cloud. P1 and D1 coincide well with each other after a displacement of 0.4 pc; the method to maximize the overlapping integral between the intensity enhancement and depression was employed by using H function defined by \citet{fuk17b}, which optimizes the projected displacement between the small cloud and the cavity created in the large cloud in a cloud-cloud collision model (Figure 7). The displacement between the small cloud and the cavity reflects a tilt angle of the relative cloud motion to the line of sight, and synthetic observations of two colliding clouds based on numerical simulations provide details of the collisional interaction between the small cloud and the cavity \citep{fuk17b}. More details are described in Section 4.2.

It could be argued alternatively that the two velocities are due to acceleration by the late O/early B stars and not a cause of cloud-cloud collision. The cloud velocity and dispersion, however, show no systematic enhancement or variation toward or correlated with the early B stars (see Figures 2), which is not consistent with a dominant dynamical effect by the stars. The velocity field which does not show particular variation toward the stars is odd, if the stars were the major source of the cloud momentum. So, we do not consider the stellar acceleration, and explore the cloud-cloud collision as a possible scenario in the present paper.

\subsection{Simulations of the collision}
In order to gain an insight into the collision, we present the numerical simulations by \citet{tak14} following \citet{fuk17b}. The simulated model is similar to that in \citet{fuk17b} except for some change in direction of the relative cloud motion which is appropriate for NGC 2024. The model deals with a head-on collision between a small cloud and a large cloud and the both are spherically symmetric. The radius of the small one is 3.5 pc and that of the large one 7.2 pc. The instantaneous collision speed is about 7 km s$^{-1}$, and the clouds have internal turbulence of 1 - 2 km s$^{-1}$ with highly inhomogeneous density distribution. Physical parameters of the collision and the clouds in these simulations are summarized in Table 1. We note that the model parameters are not finely tuned to the observations and the comparison is not so quantitative. The cloud size is larger in the model than the NGC 2024 cloud by a factor of ten, while the velocity is nearly the same between the model and observation. We consider that the difference is not a serious matter for a present comparison, which is limited mainly on patterns of the velocity field. We also point out that the cloud free fall time 5.7 Myr is significantly longer than 1.6 Myr, implying that the cloud gravity dose not play a role in the patterns.

Figure 8a shows a schematic view of the collision seen from the direction perpendicular to the cloud relative motion at two epochs, 0.0 Myr and 1.6 Myr, after the onset of the collision. The small cloud is producing a cavity in the large cloud by the collisional interaction at 1.6 Myr. The interface layer of the two clouds has enhanced density by collision, and the turbulence mixes the inhomogeneous gas of the clouds into the layer. Figures 8b-8i shows velocity-channel distributions every 0.9 km s$^{-1}$ as seen from a direction $\theta$ = 45-degree tilted to the relative motion. In the panel of Figure 8 the small cloud is seen at a velocity range from 2.0 to 4.7 km s$^{-1}$, and the large cloud from -1.5 to 1.1 km s$^{-1}$. The velocity range from 1.1 to 2.0 km s$^{-1}$  corresponds to the intermediate velocity layer. Note that the velocity ranges of each panel in Figures 8b-8i do not exactly fit the velocity ranges of the two clouds due to the turbulent mixing. The small cloud is flattened perpendicular to the traveling direction due to the merging into the intermediate layer, and the large cloud has intensity depression corresponding to the cavity created by the small cloud.

Figure 9a shows a position-velocity diagram of the model taken in the direction of the displacement, the R-axis, in the synthetic observations of the cloud-cloud collision model. In Figure 9a the large cloud is extended with an intensity depression at X $\sim$ -3 pc and the small cloud appears as a blob at X $\sim$ -2 pc. The velocity range 1 - 2 km s$^{-1}$ corresponds to the intermediate velocity layer. In Figure 9a the two clouds are not seen separately but are merged into a singe entity ranging from -1 to 6 km s$^{-1}$. Figure 9b shows the observe longitude-velocity diagram in NGC 2024. We find the peak of the small cloud at R = 0.0 - 0.2 pc and V = 10 - 12 km s$^{-1}$ and the cavity at R = 0.25 - 0.4 pc and V = 8 - 11 km s$^{-1}$. The edge of the distribution in R = -0.2 - 0.1 pc. and V = 10 - 13 km s$^{-1}$ in Figure 9b shows a gradient similar to the model in X = -6 - -4 pc and V = 1 - 4 km s$^{-1}$ in Figure 9a. These properties are generally consistent with the observed ones.

 A major qualitative difference between the model and the observation is the broad linewidth of toward the peak of the small cloud. Figure 2 shows that the linewidth toward P1 is 5 km s$^{-1}$ at the half intensity level, which is by a factor of 2 larger than that assumed in the simulations. The linewidth makes the peak elongated in velocity in Figure 9b, which is not seen in the simulation in Figure 9a. We also note the large cloud has a smaller extent limited to R $\le$ 0.7 pc, which in not taken into account in the model.

\subsection{Triggered star formation in collision}
Based on the above results and discussion we hypothesize that collision between the two components triggered formation of the late O/early B stars in NGC 2024. For simplicity, if we assume a tilt angle of 45 degrees to the line of sight, the cloud relative velocity and the displacement are 3 km s$^{-1}$ and 0.6 pc, respectively. The timescale of the collision is then estimated to be 2 $\times$ 10$^5$ yrs from a ratio 0.6 pc/3 km s$^{-1}$. This is consistent with a very small age less than Myr of the young stars in NGC2024 \citep{ali95,mey96}. Table 2 lists the cloud parameters in the present collision scenario. If we assume a tilt larger than 60 degrees the velocity becomes 4 km s$^{-1}$ or more, whereas an assumption on the tilt angle does not significantly alter the timescale. The blue-shifted component in the foreground of the H{\sc ii} region is consistent with an epoch after the collision.

Figure 10 shows the $^{13}$CO clump toward the primary peak P1; this indicates a strong concentration of 20 protostars \citep{meg12}, corresponding to an O star IRS2b and probably young B stars \citep{get14,bik03,ski03}, toward the $^{13}$CO peak P1, showing that the $^{13}$CO clump is forming high-mass stars actively.

In the present scenario, the relative velocity of the clouds is pre-determined in the Galactic environment and the collision is by chance. The gravity of the system is actually not dominant for a set of relevant parameters; the velocity in gravitational balance with an observed cloud mass of 200 $M \odot$ and a radius of 1 pc is 1.5 km s$^{-1}$, marginally less than the projection-corrected velocity above. In the scenario, the major collision took place toward the $^{13}$CO peak P1 in the red-shifted cloud, which was impacted by the blue-shifted component on the far side of the red-shifted component. The $^{13}$CO peak has a high column density of more than 10$^{23}$ cm$^{-2}$ and 20 protostars are forming at present \citep{meg12}. The interface layer between the two colliding clouds becomes highly turbulent due to the clumpy distribution in the cloud prior to the collision. The turbulence amplifies the magnetic field according to the numerical simulations of cloud-cloud collision \citep{ino13}. The combined contribution of the turbulence and magnetic field realizes high-mass accretion rate of 10$^{-3}$ to 10$^{-4}$ $M \odot$/yr, which allows a protostar to overcome the stellar radiation pressure and to grow in mass. The most massive star in NGC 2024 IRS2b having 23 $M \odot$ \citep{bik03} can be formed in a timescale of 1 $\times$ 10$^5$ yrs at an assumed mass accretion rate 2 $\times$ 10$^{-4}$ $M \odot$/yrs. The heavy obscuration toward the young stars and the sign of disks of the stars \citep{mey96} is consistent with the scenario and do not exclude that the stars are still growing in mass via accretion to become O stars eventually in $\sim$ 10$^5$ yrs, where ionization by O stars may hold further mass accretion.

\subsection{Comparison with Orion A}
\citet{fuk17b} presented a scenario that in M42 and M43 cloud-cloud collision triggered formation of the O/early B stars. The collision is supported by the complementary distributions found over 1 pc in the Orion A cloud. The collision velocity is 7 km s$^{-1}$ and the timescale is 3 $\times$ 10$^5$ yrs. The velocity is somewhat larger than that in NGC 2024, while the timescale is similar in the order of 10$^5$ yrs.

In the cloud-cloud collision model of M42 \citep{fuk17b}  the near side of the blue-shifted cloud with high column density (OMC-1 clump) collided with the red-shifted cloud. The ten O stars in M42 including Trapezium were formed on the near side of the blue-shifted cloud and the lower column-density red-shifted cloud is dissipated there by collision and ionization. The extinction Av toward these O stars is generally less than 3 magnitude, and is consistent with the low extinction expected in the cloud-cloud collision scenario \citep{sca11}. In NGC 2024 the 20 protostars were formed on the far side of the red-shifted cloud. The molecular column density toward the $^{13}$CO peak is estimated to be 1 $\times$ 10$^{23}$ cm$^{-2}$ from the $^{13}$CO ($J$=2--1) emission, which corresponds to Av $\sim$ 100 mag. Av toward IRS2b is measured to be high as 24 mag \citep{bik03}. The offset of IRC2b from the $^{13}$CO peak (Figure 10) and possible spatial variation in Av may explain in part the lower extinction than the peak value. A higher resolution CO measurement will offer more realistic column density toward the stars.

\subsection{NGC2024 in samples of cloud-cloud collision}
\citet{fuk17b} summarized observational signatures of high-mass star formation under triggering by cloud-cloud collision as follows;
\begin{enumerate}
\renewcommand{\labelenumi}{\roman{enumi})}
\item Two clouds with supersonic velocity separation associate with young high-mass star(s),  
\item complementary spatial distribution between the two clouds, and 
\item bridge feature connecting the two clouds in velocity.
\end{enumerate}

These signatures are however not always detectable in observational data. The numerical simulations (Takahiro et al. 2015, and the synthetic observations by Fukui et al. 2017b) indicate that the two clouds often show a single spectral peak instead of two peaks, because the collision mixes the two clouds in velocity as a result of momentum exchange between the two clouds (see Figures 3 and 4 in Fukui et al. 2017b). The trend of a single peak in colliding clouds becomes significant if the projected velocity separation is smaller than the linewidth of the individual clouds. The peak velocity of the merging clouds governed by the cloud with higher molecular column density (see Fukui et al. 2017\b). The collisional dissipation further destroys the two clouds \citep{tor15}, and in addition, the ionization by the formed O star(s) disperses the parent molecular gas within $\sim$ 10 pc of the O stars. As a result, the collision signatures quickly disappear (Figure 1 in Fukui et al. 2015). These effects make it difficult to identify the collision signatures and the initial two clouds, suggesting that a seemingly single cloud does not exclude cloud-cloud collision.

In NGC 2024, the projected velocity separation between the two components is small, only 2 km s$^{-1}$, and we see the complementary distribution (Section 4.1). The two components appear to be merged toward the primary peak P1 which show large velocity dispersion from 8 to 13 km s$^{-1}$ at 3 K pc level (Figure 9b). Due to the small timescale, the collisional cloud dissipation is still not significant and we see a massive dense clump toward the 20 protostars as P1 whose molecular mass is estimated to be 60 $M \odot$ from $^{13}$CO ($J$=2--1) data by assuring LTE within 0.2 pc. The absence of early O stars makes ionization less effective than in O star forming regions. In summary, the NGC 2024 cloud shows two velocity components (Figure 2b) i), and complementary distribution (Fugire 6) ii), but show no bridge feature due to the small velocity separation, satisfying two of the above conditions of cloud-cloud collision. No bridge feature is a reasonable consequence of the small velocity separation of the two clouds and is not against cloud-cloud collision.

On high-mass star formation, the nine cases where O stars were formed by cloud-cloud collision suggest that there is a threshold column density for forming O stars(s) \citep{fur09,oha10,tor11,tor15,tor16,tor17,fuk14,fuk15,fuk17a,fuk17b,fuk18b}.
 The threshold molecular column density in formation of multiple O stars is $\sim$ 10$^{23}$ cm$^{-2}$ and that in a single O star 10$^{-22}$ cm$^{-2}$. For column density less than these values no O stars are formed in cloud-cloud collision, and only collision toward high column-density gas leads to formation of O/early B stars. The high column density in NGC 2024, $\sim$ 10$^{23}$ cm$^{-2}$, is consistent with the threshold, and we infer that even more massive stars may form in near future if high-mass star formation continuing in NGC 2024. A further discussion of cloud-cloud collision and high-mass star formation in Orion will be developed in a forth coming paper on NGC 2068 and NGC 2071 \citep{tsu18}

\section{Conclusions}
We carried out new observations of $^{13}$CO ($J$=2--1) transitions in the NGC 2024 cloud with NANTEN2. These observations cover the whole NGC 2024 cloud at 0.2 pc resolution. The main results are summarized below. 

\begin{enumerate}
\item Contrary to the previous observations which suggested a single cloud component, we found a possibility that the cloud comprises two velocity components with complementary distribution. The projected velocity separation of these clouds is $\sim$ 2 km s$^{-1}$. The blue-shifted component shows good correspondence with several minor dark lanes, indicating that it is on the foreground of the H{\sc ii} region. 

\item We found that a displacement of 0.4 pc in a position angle of 60 degrees produces a good spatial correspondence between the intensity peaks and depressions forming complementary distribution. The two velocity components seem to be merged into a single velocity component of $\sim$ 4 km s$^{-1}$ velocity span at a 0.1 K deg level (Figure 8a).

\item We hypothesize that collision between the two clouds at 9.5 km s$^{-1}$ and 11.5 km s$^{-1}$ triggered formation of the 20 protostars including IRS2b possibly ionizing NGC 2024. The collision timescale is estimated to be 2 $\times$ 10$^{5}$ yrs. The molecular column density toward the $^{13}$CO peak which is not self-absorbed is estimated to be $\sim$ 10$^{23}$ cm$^{-2}$, and that in the surrounding regions is $\sim$ 10$^{22}$ cm$^{-2}$. The collision realized a high-mass accretion rate of 10$^{-4}$ $M \odot$/yrs as shown by the mechanism presented by \citet{ino13}, triggered the O/early B star formation.
\end{enumerate}

It is important in future to investigate nearby molecular clouds within a few kpc in order to better establish the role of collision in O star formation.
The coarse resolution CO data in Figure 1 does not exclude two velocity components, which might suggest a cloud-cloud collision toward NGC2023. It is therefore important to explore a possibility of cloud-cloud collision at higher resolution toward NGC 2023 also. NGC 2024 is a unique object which is extremely young with an age less than $10^5$ yrs among those discovered until now. Thanks to the young age the parent cloud remains unionized and we are able to observe its details including the column density. Higher resolution studies with ALMA will shed a new light on the formation of dense clumps and their mass function in the shock-compressed layer by collision.

\begin{ack}
This work was financially supported Grants-in-Aid for Scientific Research (KAKENHI) of the Japanese society for the Promotion for Science (JSPS; grant number society for 15K17607 and 15H05694). NANTEN2 is an international collaboration of ten universities, Nagoya University, Osaka Prefecture University, University of Cologne, University of Bonn, Seoul National University, University of Chile, University of New South Wales, Macquarie University, University of Sydney, and Zurich Technical University.
\end{ack}



\newpage
\begin{table}[h] 
\begin{center}
\tbl{The initial conditions of the numerical simulations (Takahira et al. 2014) }{%
\begin{tabular}{@{}cccccc@{}} \noalign{\vskip3pt}
\hline\hline
\multicolumn{1}{c}{Box size [pc]} & $30 \times 30 \times 30$  &  &  &  \\ [2pt]
\noalign{\vskip3pt}
Resolution [pc] & 0.06 &  &  \\
Collsion velocity [km s$^{-1}$] & 10 (7)$^{\dag}$ & & \\
\hline
Parameter & The small cloud & The large cloud & note    \\
\hline
Temperature [K] & 120 & 240 &  \\
Free-fall time [Myr] & 5.31 & 7.29 &  \\
Radius [pc] & 3.5 & 7.2 &  \\
Mass [$M_{\odot}$] & 417 & 1635 &  \\
Velocity dispersion [km s$^{-1}$] & 1.25 & 1.71 &  \\
Average Density [cm $^{-3}$] & 47.4 & 25.3 & Assumed a Bonner-Ebert sphere  \\
\hline\noalign{\vskip3pt} 
\end{tabular}} 
\label{tab:first} 
\begin{tabnote}
{\hbox to 0pt{\parbox{150mm}{\footnotesize
Note. \footnotemark[$\dag$] The initial relative velocity between the two clouds is set to 10 km s$^{-1}$, whereas the collisional interaction decelerates the relative velocity to about 7 km s$^{-1}$ in 1.6 Myrs after the onset of the collision. The present synthetic observations are made for a relative velocity 7 km s$^{-1}$ at 1.6 Myrs.
\par\noindent
\phantom{0}
\par
\hangindent6pt\noindent
}\hss}}
\end{tabnote} 
\end{center} 
\end{table}

\newpage
\begin{longtable}{*{8}{l}}
\caption{The model parameters of the two clouds in NGC 2024}
\multicolumn{8}{c}{} \\
    Parameter & The blue-shifted cloud & The red-shifted cloud \\
  \hline
\endfirsthead
\hline
\hline
\endhead
\hline
\endfoot
\hline
\multicolumn{8}{l}{} \\
\endlastfoot
    Velocity range\footnotemark[$\dag$$_1$] [km s$^{-1}$] & 7.0\,-\,10.5 & 10.5\,-\,14.7 \\
    Length\footnotemark[$\dag$$_2$] [pc] & 1.0 & 1.0 & \\
    Width\footnotemark[$\dag$$_2$] [pc] & 0.5 & 0.5 & \\
    Mass\footnotemark[$\dag$$_3$] [$M_\odot$] & 5.4 $\times$ 10$^{2}$ & 3.6 $\times$ 10$^{2}$  \\
\end{longtable}
\begin{tabnote}
\hangindent8pt\noindent
 Note. \footnotemark[$\dag$$_1$]The velocity range between the blue-shifted cloud and the red-shifted cloud decided the coincidence with dark lanes as shown in Figure 4g. \footnotemark[$\dag$$_2$]The length and width are the size within half intensity of the peak integrated intensity. \footnotemark[$\dag$$_3$]The molecular mass is estimated from $^{13}$CO($J$=2--1) data by assuring LTE, where the excitation temperature 
 is assumed to be 20 K.
\end{tabnote}

\newpage
\begin{figure*}
\begin{center}
 \includegraphics[width=8cm]{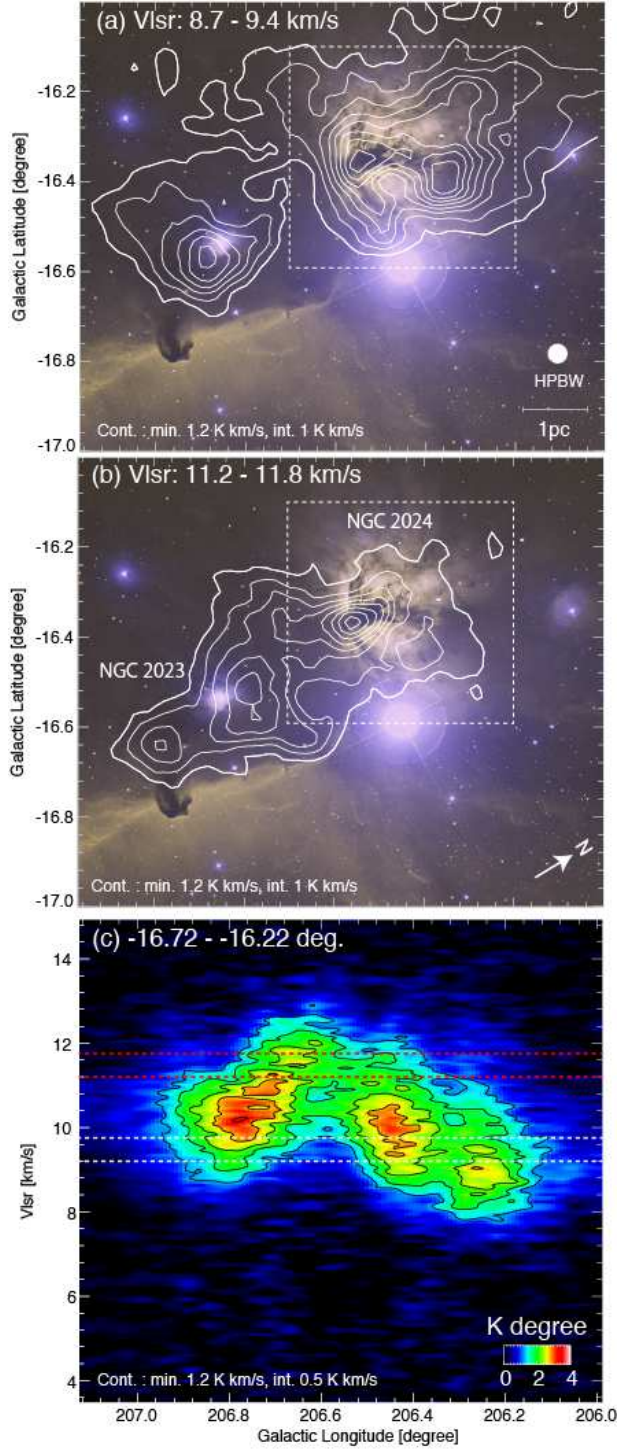}
   \end{center}
\caption{Composite of two optical images overlaid on the 1.85 m telescope $^{13}$CO($J$= 2--1) integrated intensity contours toward NGC~2023 and NGC~2024 (blue: DSS2 blue, red: DSS2 red) (STScI Digitized Sky Survey, (c) 1993, 1994, AURA, Inc.). (a) The contour map of the blue-shifted cloud at 8.7 -- 9.4 km s$^{-1}$ are super imposed on the optical image. (b) The contour map of the red-shifted cloud at 11.2 -- 11.8 km s$^{-1}$ are super imposed on the optical image. These contours are from 1.2 K km s$^{-1}$ to 8.2 K km s$^{-1}$ every 1 K km s$^{-1}$. The area indicated by dashed-box was observed with NANTN2. (c) Position-velocity diagram of $^{13}$CO($J$=2-1) emission toward NGC 2023 and NGC 2024. The horizontal dashed lines show the velocity range for the red-shifted cloud [ 8.7 - 9.4 km s$^{-1}$] and the blue-shifted cloud [ 11.2 - 11.8 km s$^{-1}$].}
\end{figure*}

\begin{figure*}
 \begin{center}
  \includegraphics[width=17cm]{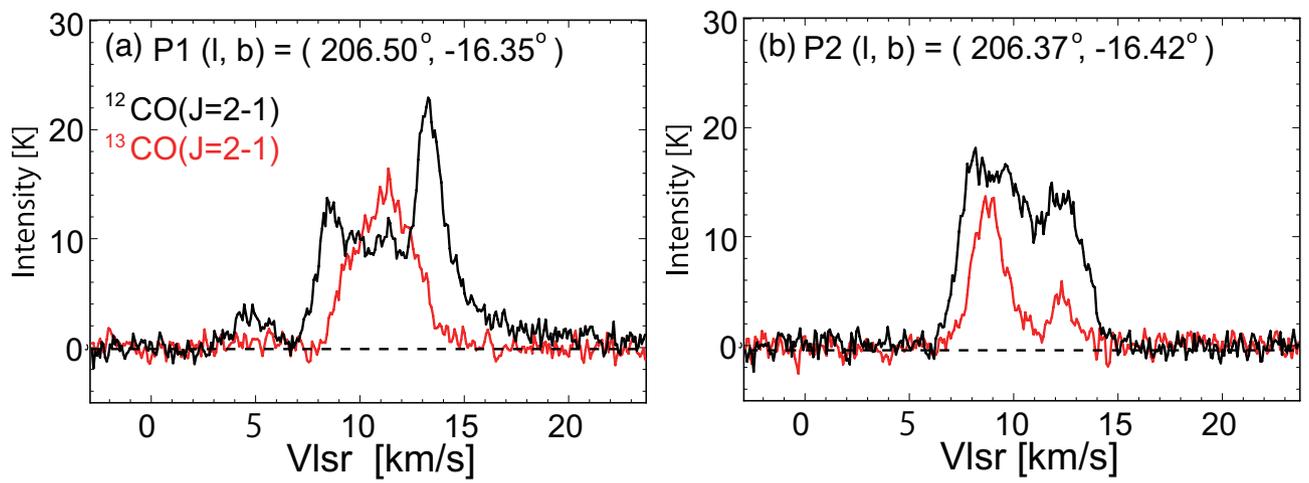} 
 \end{center}
\caption{CO($J$=2--1) spectra at two peak positions of the blue-shifted cloud and red-shifted cloud. The horizontal and vertical axis indicate a radial velocity and main-beam temperature. $^{12}$CO($J$= 2--1) and $^{13}$CO($J$= 2--1) are plotted in black and red lines, respectively.}
\end{figure*}

\begin{figure*}
 \begin{center}
  \includegraphics[width=17cm]{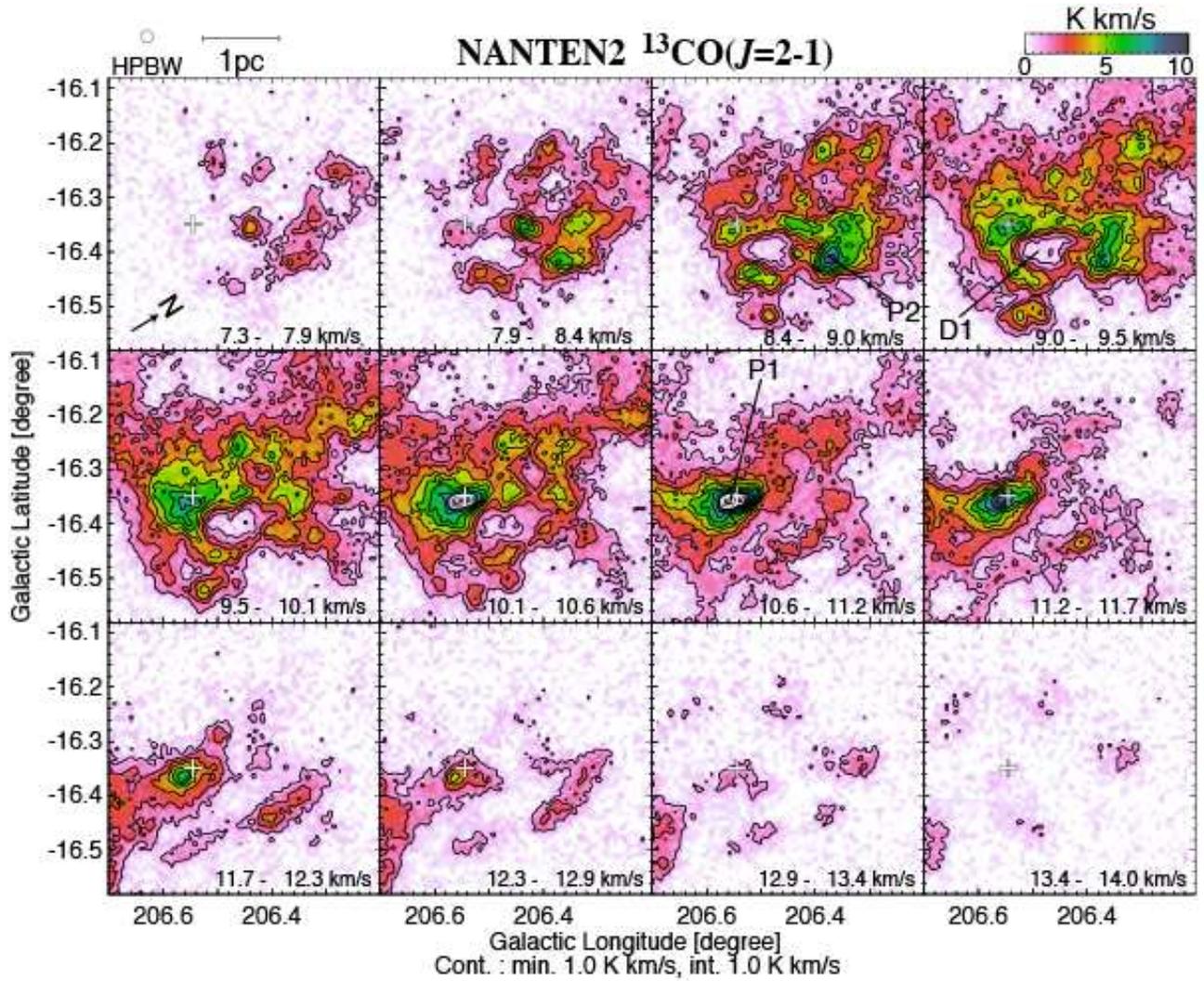} 
 \end{center}
\caption{Velocity-channel distributions of the $^{13}$CO($J$=2--1) emission toward NGC 2024 at velocity step of 0.6 km s$^{-1}$. These contours are from 1 K km s$^{-1}$ to 11 K km s$^{-1}$ every 1 K km s$^{-1}$. The cross depicts the position of IRS2b toward NGC2024. P1, P2 and D1 show the primary peak intensity ($l$, $b$)=(206.50{$^{\circ}$}, -16.35{$^{\circ}$}), secondary peak intensity ($l$, $b$)=(206.37{$^{\circ}$}, -16.42{$^{\circ}$}) and intensity depression ($l$, $b$)=(206.46{$^{\circ}$}, -16.40{$^{\circ}$}).}
\end{figure*}

\begin{figure*}
 \begin{center}
  \includegraphics[width=16cm]{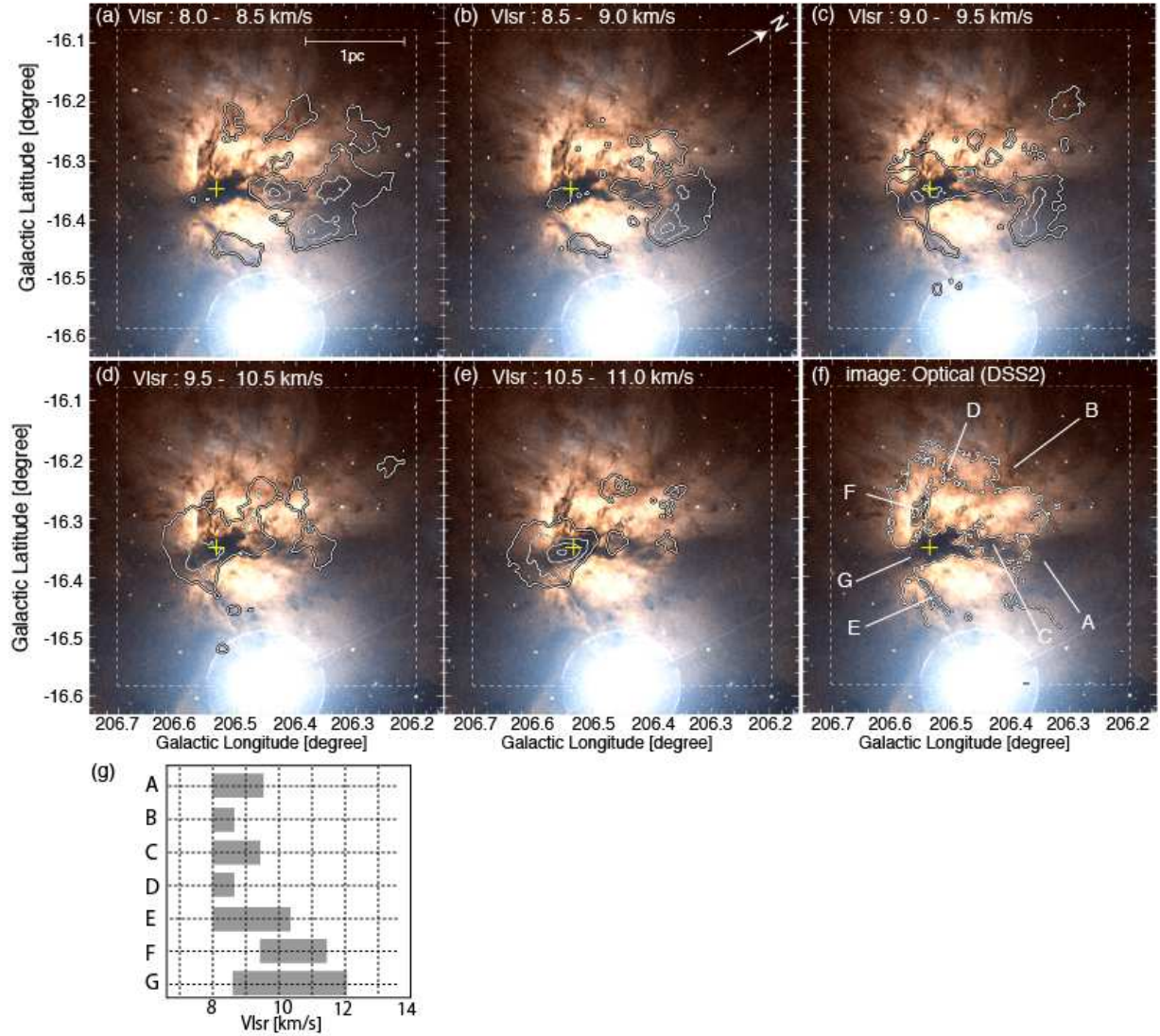} 
 \end{center}
\caption{$^{13}$CO($J$= 2--1) integrated intensity contours overlaid on optical image of NGC~2024. The panels show CO data in the velocity range of 8.0 to 8.5 km s$^{-1}$ (a), 8.5 to 9.0 km s$^{-1}$ (b), 9.0 to 9.5 km s$^{-1}$ (c), 9.5 to 10.5 km s$^{-1}$ (d), 10.5 to 11.0 km s$^{-1}$ (e) and the panel (f) optical image (blue: DSS2 blue, red: DSS2 red). These contours are from 2.0 K km s$^{-1}$ to 8.0 K km s$^{-1}$ every 2.0 K km s$^{-1}$ in these panels (a-e). (f) These solid lines indicate the dark lanes. The level of white lines is 12,000 counts. Panel (g) shows the velocity range of $^{13}$CO($J$=2--1) to correspond with these dark lanes A-G.}
\end{figure*}

\begin{figure*}
 \begin{center}
  \includegraphics[width=17cm]{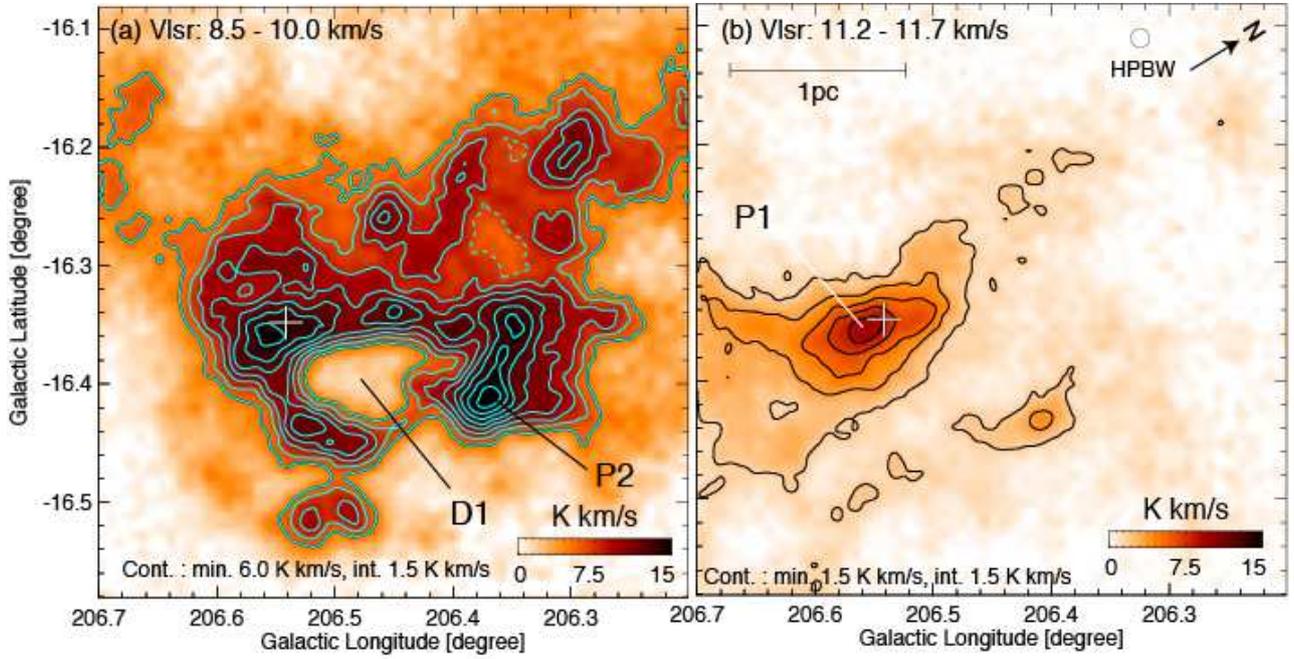} 
 \end{center}
\caption{The typical CO($J$=2--1) distributions of the blue-shifted and red-shifted clouds toward NGC 2024, where the velocities of the image and contours are 8.5 - 10.0 km s$^{-1}$ for the panel (a) and 11.2 - 11.7 km s$^{-1}$ for the panel (b). The cross denotes IRS2b as in Figure 4. P1, P2 and D1 show the primary peak intensity, secondary peak intensity, and intensity depression.}
\end{figure*}

\begin{figure*}
 \begin{center}
  \includegraphics[width=17cm]{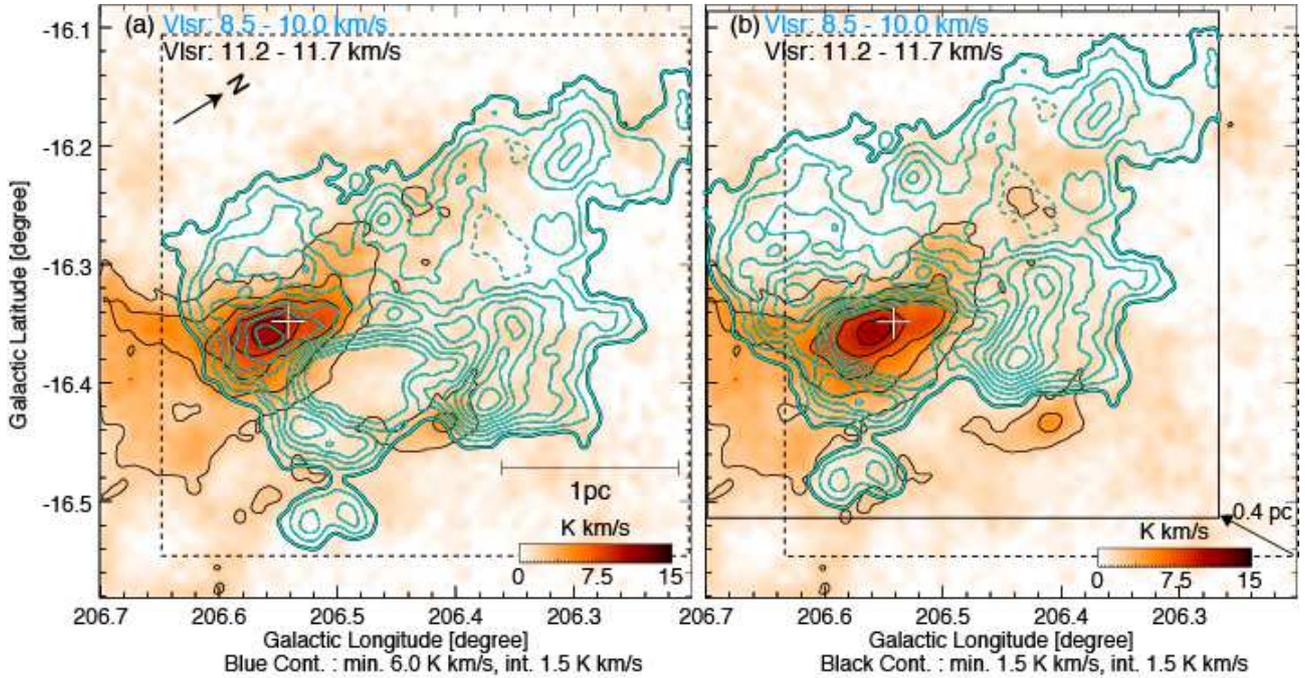} 
 \end{center}
\caption{The region of the complementary distribution in the box with dashed lines. Panel (a) indicates image of Figure 5(b) overlaid with the contours of Figure 5(a). The image and the black contours show the red-shifted cloud (11.2 - 11.7 km s$^{-1}$) and the blue contours show the blue-shifted cloud (8.5 - 11.0 km s$^{-1}$). The white cross denotes the IRS2b of NGC 2024. Panel (b) indicates the complementary distributions of the two velocity distributions after the displacement 0.4 pc shown by the arrow of 60-degree angle. The image with black contours and blue contours indicate the red-shifted cloud and blue-shifted cloud, respectively. The blue and black contours are from 6.0 K km s$^{-1}$ every 1.5 K km s$^{-1}$ and 1.5 K km s$^{-1}$ every 1.5 K km s$^{-1}$, respectively.}
\end{figure*}

\begin{figure*}
 \begin{center}
  \includegraphics[width=17cm]{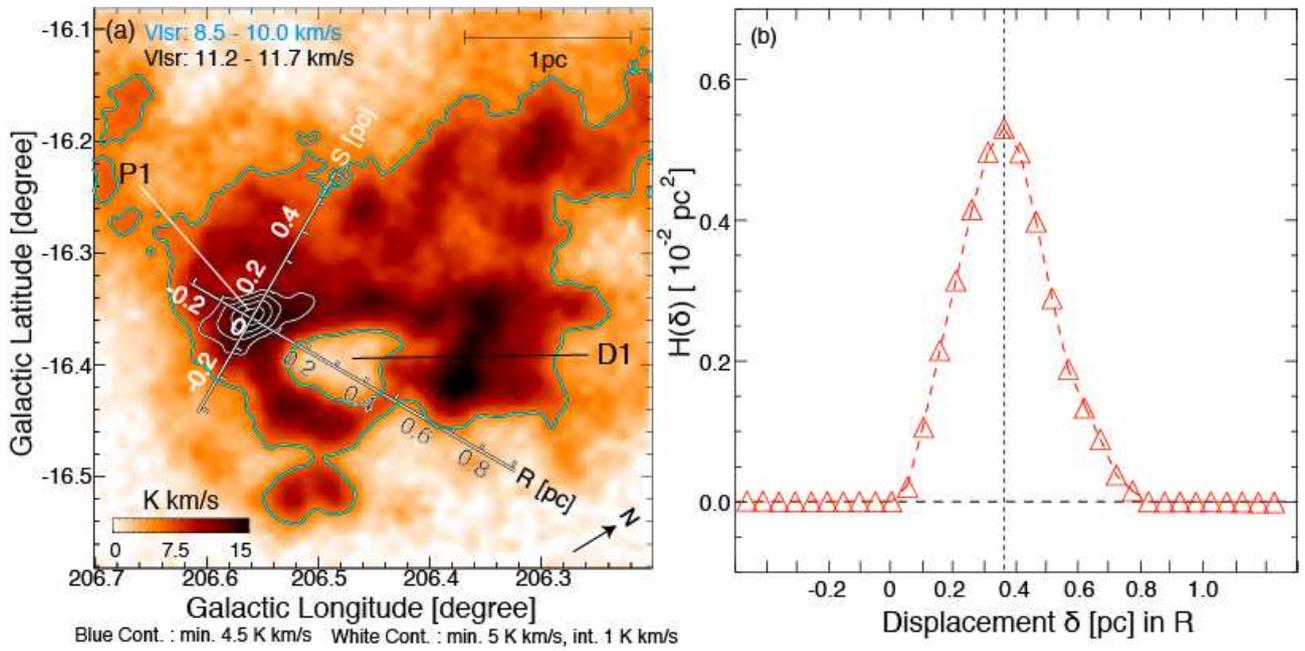} 
 \end{center}
\caption{(a) The image and the blue contours show the blue-shifted cloud including the intensity depression D1, and the white contours show the red-shifted cloud including the peak intensity P1. The blue and white contours indicate the lowest intensity limit 3 and 5 K km s$^{-1}$. R-axis was taken for the direction of the displacement and the S-axis is normal to the R-axis. Panel (b) indicates the overlapping integral H($\delta$) for the peak intensity P1 and depression intensity D1, where the peak intensity P1 and intensity depression D1 ere assumed to be a uniform value 1.0 (arbitrary unit).}
\end{figure*}

\begin{figure*}
 \begin{center}
  \includegraphics[width=16cm]{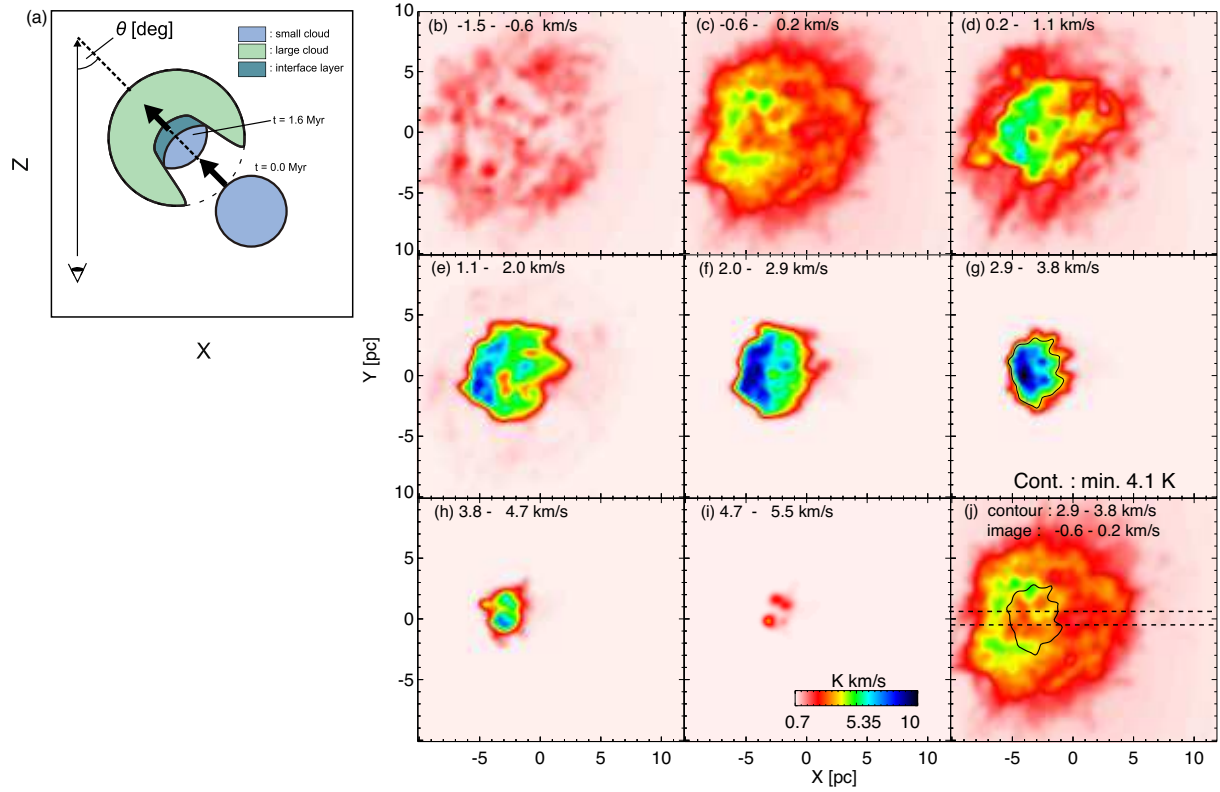} 
 \end{center}
\caption{The numerical model by Takahira et al. (2015) at 1.6 Myr observed at an angle of 45 degrees between the line of sight and the cloud relative motion. Panel (a) is a schematic of the top-view of the collision. Panels (b-i) show the velocity channel distributions every 0.9 km s$^{-1}$ in a velocity interval indicated in each panel. Panel (j) shows an overlay between the large cloud, the image in Panel 8(c), and the small cloud with the contour of Panel 8(g). The black contour is that of 4.1 km s$^{-1}$ in panel (g).}
\end{figure*}

\begin{figure*}
 \begin{center}
  \includegraphics[width=12cm]{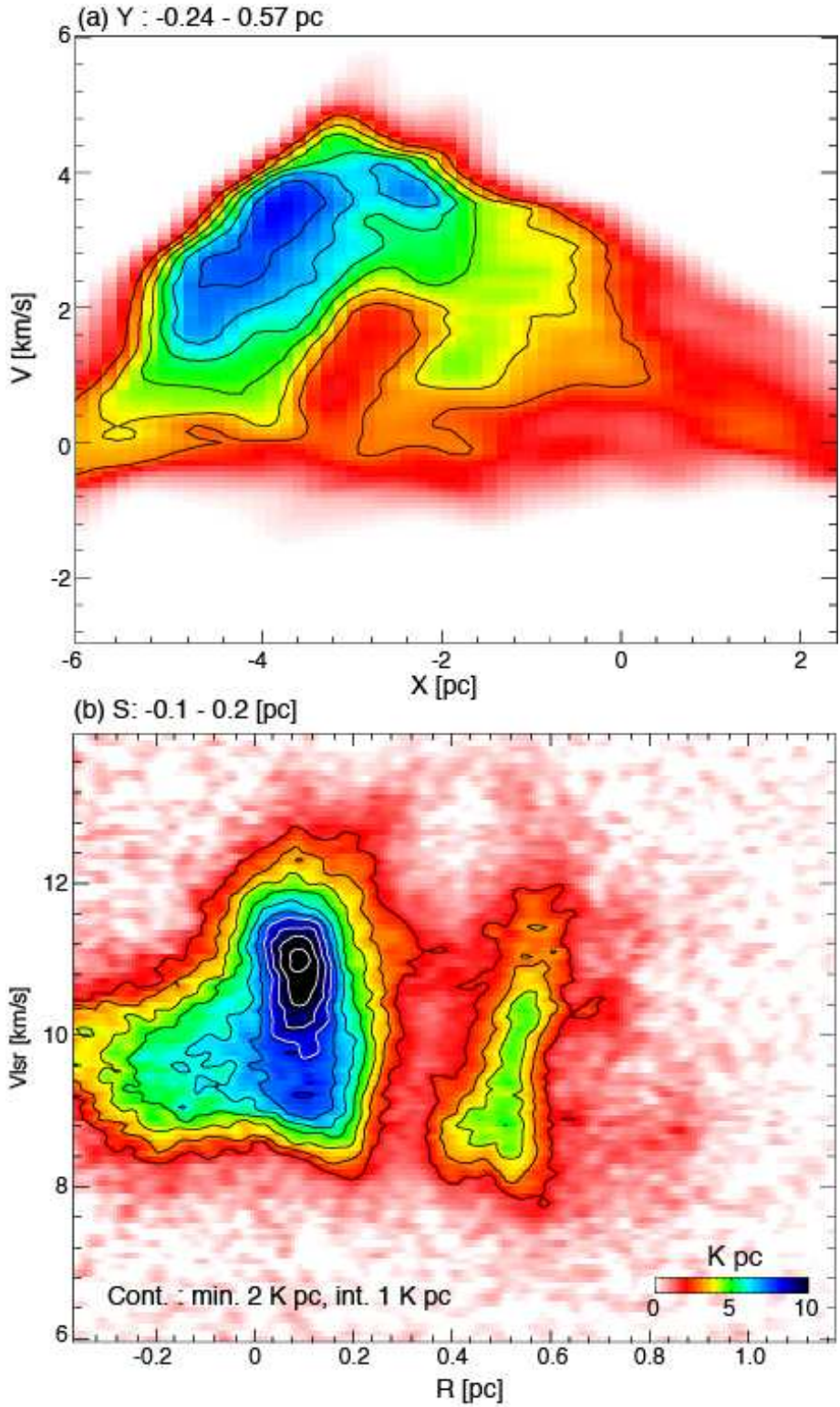} 
 \end{center}
\caption{Panel (a) shows a longitude-velocity diagram of the cloud-cloud collision at 1.6 Myr based on the numerical simulations by Takahira et al. (2015). The integration range is shown Figure 8(j) by dashed lines. The contour level is from 4 K pc to 8 K pc every 1 K pc. Panel (b) shows the observed position-velocity diagram of $^{13}$CO($J$=2--1). The integration range is from -0.1 pc to 0.2 pc for S-axis. The contour level is from 2 K pc to 11 K pc every 1 K pc.}
\end{figure*}

\begin{figure*}
 \begin{center}
  \includegraphics[width=15cm]{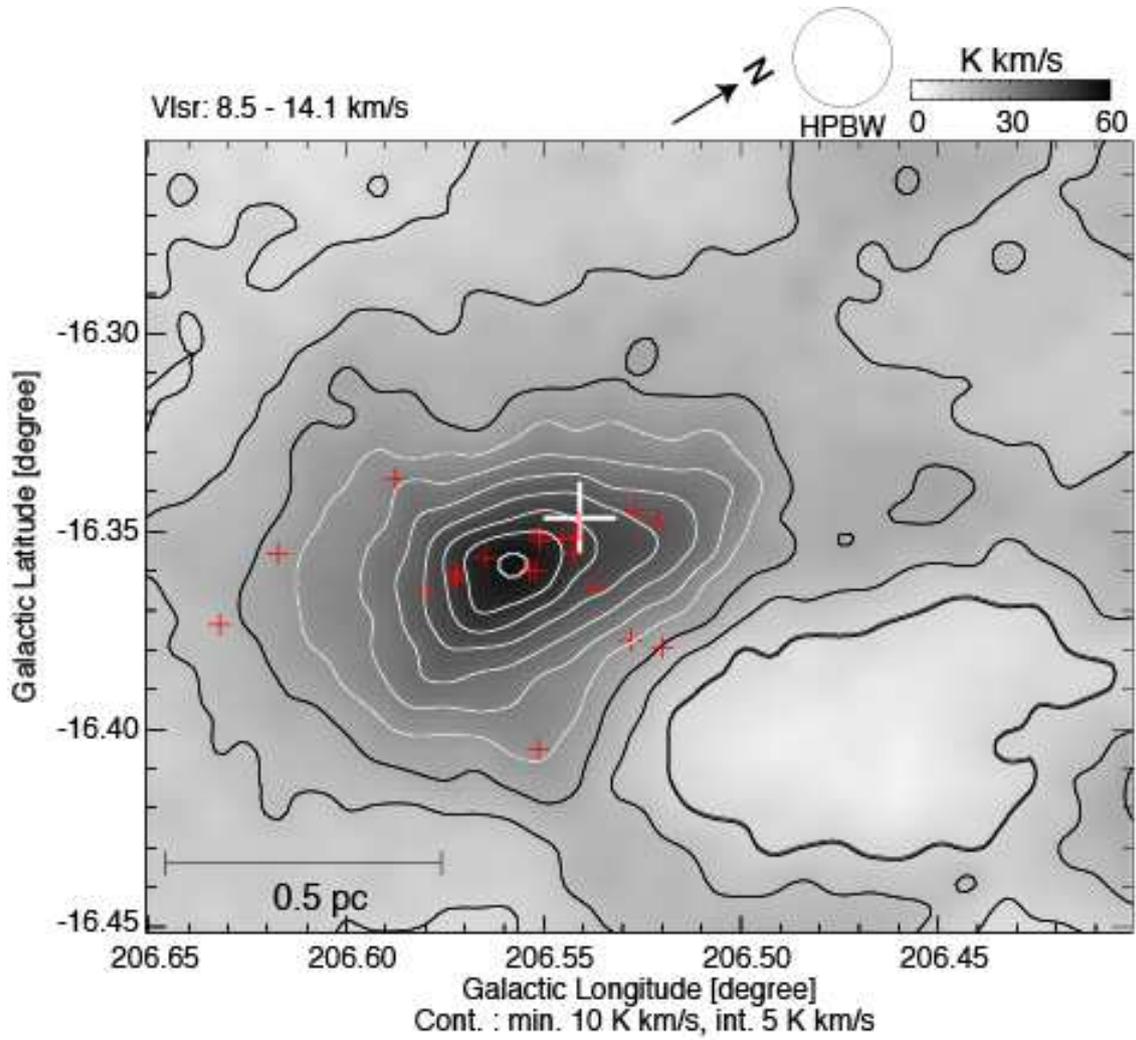} 
 \end{center}
\caption{The $^{13}$CO($J$=2--1) distributions of the red-shifted clouds toward the NGC 2024, where the velocities of the contours an the image are  8.5 -- 14.1 km s$^{-1}$. The contour level is from 10 K km s$^{-1}$ to 55 K km s$^{-1}$ every 5 K km s$^{-1}$. The white cross and the red crosses denote IRS 2b (Bik et al.(2003)) and protostars (Megeath et al.(2012)).}
\end{figure*}

\end{document}